\documentclass[12pt,preprint]{aastex}

\usepackage{emulateapj5}
\renewcommand\plotone[1]{
 \centering \leavevmode
 \columnwidth=1.55\columnwidth 
 \includegraphics[width={\columnwidth}]{#1}
}
\renewcommand\plottwo[2]{%
 \centering \leavevmode
 \columnwidth=.72\columnwidth
 \includegraphics[width={\columnwidth}]{#1} \hfil 
 \includegraphics[width={\columnwidth}]{#2}
}

\newcommand{\Alfven}{Alfv\'{e}n }

\slugcomment{Received 2001 September 18; accepted 2001 October 9; published 2001 November 6}

\shorttitle{Particle Acceleration in Relativistic Magnetic Reconnection}
\shortauthors{Zenitani and Hoshino}

\begin{document}

\title{The Generation of Non-thermal Particles in \\
  Relativistic Magnetic Reconnection of Pair Plasmas}

\author{S. Zenitani and M. Hoshino}

\affil{Department of Earth and Planetary Science, University of Tokyo,
7-3-1, Hongo, Bunkyo-ku, Tokyo, 113-0033, Japan.}

\begin{abstract}
Particle acceleration in magnetic reconnection of
electron-positron plasmas is studied
by using a particle-in-cell simulation.
It is found that significantly large number of non-thermal particles
are generated by the inductive electric fields
around an X-type neutral line
when the reconnection outflow velocity,
which is known to be an \Alfven velocity,
is of the order of the speed of light.
In such a relativistic reconnection regime,
we also find that electrons and positrons
form a power-law-like energy distribution
through their drift along the reconnection electric field
under the relativistic Speiser motion.
A brief discussion of the relevance of these results
to the current sheet structure, which has an anti-parallel magnetic field
in astrophysical sources of synchrotron radiation, is presented.
\end{abstract}


\singlespace

\section{Introduction}

The non-thermal particles in space plasmas
have attracted our attention for a long time.
Their origins remain central problems and
many kinds of acceleration processes have been studied.
Double layer and shock acceleration mechanism are important ones.

Magnetic reconnection is one of the fundamental processes in plasmas.
It is well accepted that it plays a crucial role
in the Earth's magnetotail, solar corona and astronomical accretion disks.
Since the stored magnetic energy is rapidly released to
particle kinetic energy,
a reconnection process is also important
as one of the acceleration mechanisms in plasmas.
In astronomical high-energy situations,
several radio observations \citep{lesch92,dejager94}
have suggested that magnetic reconnection
may be the source of non-thermal radiations.
Magnetic reconnection in pair-plasmas is also discussed
for energetics of a striped wind of the Crab pulsar
\citep{michel82,coro90,michel94,lyu01}.

According to the studies on reconnection of the Earth's magnetotail,
particle acceleration in reconnection mainly takes place
around the X-type neutral region,
where the amplitude of magnetic fields is weak
and charged particles become unmagnetized.
They are accelerated by an inductive electric field
perpendicular to the two-dimensional reconnection magnetic fields.
Based on the studies on particle behavior
around the X-type region \citep{Speiser65,Sonnerup},
many authors \citep{zele90,Deeg91} have investigated
the particle energetization by test particle simulations,
in the time-stationary/dependent
electric and magnetic fields obtained by MHD simulations.
They have demonstrated that accelerated particles
form power-law-like energy spectra.

In order to study the acceleration in reconnection fields more precisely,
we should consider self-consistency
between particle motion and electromagnetic fields.
Thus a full-particle simulation
that follows the kinetic plasma equations is required.
However, many of the full-particle studies of reconnection
have focused on field structure or energy conversion.
The acceleration of energetic particles in reconnection
is not discussed enough in spite of its importance.
\citet{hoshi_supra} have discussed supra-thermal electrons accelerated
near the magnetic pile-up region, in addition to the X-type region.

\begin{figure*}[thp]
\plotone{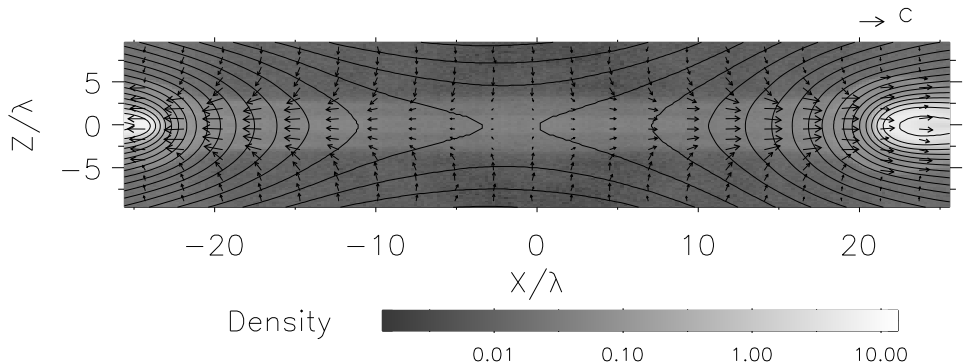}
\caption{A snapshot of magnetic reconnection region
of $-25.6 \le X/\lambda \le 25.6$ and $-9.6 \le Z/\lambda \le 9.6$
at $t / \tau_A = 80.6$.
The solid lines and the vectors show
magnetic field lines and the plasma flows, respectively.
The color contour represents the plasma density,
which is normalized by the initial density in the plasma sheet $n_0$.
\label{rec}}
\end{figure*}

In this letter, we study particle acceleration
in magnetic reconnection of astronomical pair-plasmas,
using PIC (particle-in-cell) code.
We chose the condition that
typical \Alfven velocity of plasma is the order of light speed.
From our simulation results, we have found
remarkable amount of non-thermal component in the energy spectrum
and found that this is a sign of acceleration
by the strong inductive electric field near the X-type region.
The high-energy particles are accelerated
through Speiser/meandering-like orbit around there,
and their maximum energy has grown up to $mc^2\Omega_cT$,
where $\Omega_c$ is the cyclotron frequency,
and $T$ is the typical reconnection time which may be defined by $L/c$,
where $L$ is the size of the reconnection region.
Moreover, the spectra around the X-type region
seem to have a power-law distribution with the power-law index of 1.
We consider that
feedback effect of relativistic inertia plays an important role
in formation of the spectra, 
and propose a new process of formation of this spectrum.

This letter is organized as follows.
In the next section we present our simulation conditions.
In Sec. 3, we show several results and findings of our run.
In Sec. 4, we study the acceleration in the vicinity of the X-type neutral point
and introduce basic idea of formation of power-law spectrum.
In Sec. 5, we summarize and conclude our study.

\section{Simulation model}

We use a high-resolution relativistic electro\-magnetic PIC (particle-in-cell) code.
The evolution of two-dimensional electromagnetic configuration is considered.
We calculate the all three components of particles' positions and velocities,
and observe the field structure in the X-Z simulation plane.
All quantities are uniform in the Y direction.
For simplicity, we neglect any collisions, 
pair-production and pair-annihilation of pair-plasmas.

We study slab geometry of the plasma sheet and
started from a Harris equilibrium model \citep{Harris},
which is commonly used for the reconnection problem.
Since the standard Harris equilibrium is applicable to
only non-relativistic plasma sheet,
we extended it into the relativistic plasmas
by replacing the velocity $v$ to
the four velocity $u = v \gamma = v / [ 1-(v/c)^2 ]^{1/2}$,
where $c$ is the speed of light.

The simulation region consists of $1024 \times 512$ numerical meshes
and the thickness of the plasma sheet $\lambda$ is set to 10 grids.
The magnetic field, the plasma density,
and distribution function of plasmas are described by
$\mathbf{B}(z) = B_0 \tanh (z/\lambda) \cdot \hat{x}$,
$n(z) = n_0 \cosh^{-2}(z/\lambda)$
and
$f = n(z) \exp \{ - m [ u_x^2 + (u_y-U)^2 + u_z^2 ] /{2T} \}$, 
respectively.
The typical particle kinetic energy is $ 0.25 mc^{2} $ in our condition.
The total number of particles is $6.7 \times 10^7$.
The particle density in the plasma sheet is
$n_{PS} \sim 7.7 \times 10^2$ pairs per grid,
while $n_{Lobe} \simeq 6 - 7$ pairs in the lobe.
We use double periodic boundary condition,
therefore the system size of each plasma sheet
is $-51.2 \le X/\lambda \le 51.2$ and $-12.8 \le Z/\lambda \le 12.8$.
We assume a thin plasma sheet where the thickness
is comparable with the typical Larmor radius of particles:
$\lambda = 2r_L$.

We assume that the cyclotron frequency in the lobe
is equal to the plasma frequency in the current sheet
$\Omega_c = \omega_{pe}$, where $\Omega_c = eB_0/{mc}$
and $\omega_{pe} = [4\pi n_0 e^2/m]^{1/2}$.
Thus the reconnection outflow,
whose speed is known to be an \Alfven velocity of the system
$V_A \sim c/[ 1+2(\omega_{pe} / \Omega_c)^2 ]^{1/2}$,
is expected to be the order of the speed of light.

In the very early stage of reconnection,
we drive small external electric fields
localized on the outside of the plasma sheet
in order to trigger an X-type neutral line
around the center of the simulation box.
The system slightly gains energy from these additional fields.
After the electric fields are eliminated,
we confirmed that the total energy is conserved within 0.1\% error
throughout the simulation run.

\begin{figure*}[htbp]
\plottwo{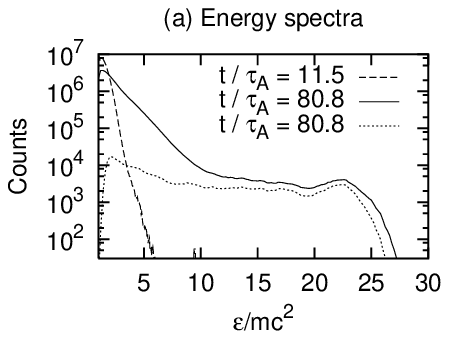}{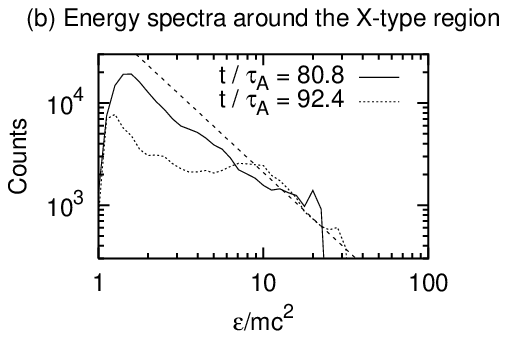}
\caption{The energy spectra in the system.
The dashed line and solid line in left panel
represent energy spectra over the whole simulation box
at $t/\tau_A = 11.5$ and $t/\tau_A = 80.6$, respectively.
The dotted line shows the partial energy of particles
only around the X-type region
($-16.0 \le X/\lambda \le 16.0$ and $-6.4 \le Z/\lambda \le 6.4$)
at $t/\tau_A = 80.6$.
The right panel is another view of the energy spectra around the X-type region
in the log-log scaling.
(solid line; $-16.0 \le X/\lambda \le 16.0$ and $-6.4 \le Z/\lambda \le 6.4$).
It is noticeable that these spectra well match
the power-law distribution with the index of 1(dashed line).
Later in the simulation time, at $t/\tau_A = 92.4$,
this spectrum evolves into the dotted line
and its highest energy edge reaches to $38 mc^2$.
It still keeps the power-law relation.
}
\end{figure*}

\section{Results}

Figure 1 shows the snapshot of the magnetic field lines and the density
at $t/\tau_A = 80.6$, where $\tau_A = \lambda / V_{A}$ (\Alfven transit time).
An X-type neutral line is formed at the center of the simulation box
and plasmas are streaming out from the X-type region
toward the $\pm$ X directions.
The maximum outflow speed reaches up to $0.91c$,
which exceeds the typical \Alfven speed in the system.
The basic behavior of the nonlinear evolution of the plasma sheet
is same as that of other MHD, hybrid, and particle simulation results
performed in non-relativistic regime.
The magnetic reconnection rate $(c E_y /B_0)/V_{out}$ is about $0.33$.
As the time goes on,
the thickness of the plasma jet becomes $\sim 2 \lambda$,
which is of the order of the meandering width of accelerated particles,
while the meandering width before reconnection was
$(\lambda r_L)^{1/2}\sim 0.7 \lambda $.

Let us study the plasma heating and acceleration
during the relativistic magnetic reconnection.
Figure 2a shows the energy spectra in the whole simulation box
at two different stages of our simulation.
In the initial growth phase ($t/\tau_A = 11.5$),
the spectrum is well described by a Maxwellian,
$f(\varepsilon) \propto \exp(-\varepsilon/T) $,
where $T \sim 0.4mc^{2}$ is the effective temperature.
As the time goes on, we can observe not only hot plasma
but also a non-thermal high-energy tail in the spectrum.
The dashed line shows the energy spectrum at $t/\tau_A = 80.8$.
One can observe a significant enhancement in the high-energy part,
and the maximum energy reaches up to $\sim 27 mc^2$.
To analyze the acceleration site of the non-thermal particle,
we show the energy spectra integrated particles
only around the X-type region of $-16.0 \le X/\lambda \le 16.0$
and $-6.4 \le Z/\lambda \le 6.4$.
The dotted line in Figure 2a indicates the above partial energy spectrum.
We find that most of the high-energy particles in the system
are produced around the X-type region.

Figure 2b shows two energy spectra around the X-type region
in the log-log scales at $t/\tau_A = 80.8$ and $92.4$.
This non-thermal part may be approximated by power-law distribution
$f(\varepsilon) \propto \varepsilon^{-s} $ with $s \sim 1$.
We also find that the maximum energy $\varepsilon_{max}$
in the simulation box constantly grows in time
and the growth is well described by $\varepsilon_{max} \sim e E_0c (t-t_0)$,
where $e$, $c$ and $E_0$ are the electric charge,
the speed of light, and the electric field around the X-type region, respectively.
The reconnection electric field $E_0$ is also found to be
almost constant during the nonlinear evolution of reconnection,
and $E_0$ is about $0.3 B_0$.
The parameter $t_0$ is the onset time of reconnection,
which is controlled by the driven electric field in the outer plasma sheet.
In this case, $t_0/\tau_A \sim 40$.
We stop the simulation at $t/\tau_A=92.4$
when the outflow plasma starts to collide in the periodic boundary.
Due to this plasma compression effect,
the growth of $\varepsilon_{max}$ ceases at $t/\tau_A \sim 92.4$,
and $\varepsilon_{max}$ finally reaches to $38 mc^{2}$ in the typical run.

In order to study how the particles gain their energy,
we first picked up some typical high-energy particles in the outflow region,
and traced their positions backward in time.
As a result, we found that most of the non-thermal particles come
through the X-type region.
The particles are initially situated on the both side of the plasma sheet,
and they are successively transported into the X-type region
as the reconnection evolves.
Once they come into the X-type region,
they are strongly accelerated along the reconnection electric field $E_y$,
and their motions are basically described by
the so-called Speiser/meandering orbit.
Due to the reconnecting magnetic field $B_z$,
the particle momentum $P_y$ are transformed to $P_x$,
and then they are ejected toward the $\pm$X directions.

\section{Acceleration around the X-type region}

Next, let us study the acceleration process
around the X-type region in more detail.
In case of the non-relativistic reconnection,
particles are accelerated by drifting
toward the reconnection electric field $E_y$
during Speiser/meandering orbit \citep{Speiser65}
around the X-type region.
The typical acceleration time scale is
described by the reciprocal of the cyclotron-frequency
determined by the reconnection magnetic field $B_z$, i.e., $ mc/(eB_z)$.

However, in our relativistic reconnection case,
the electric field $E_{y}$ is strong enough to
drive the particles into the Y direction,
and the particle energy reaches up to $mc^2$.
In this regime, the cyclotron-frequency is a function of the particle energy
$\varepsilon$
and the cyclotron-radius becomes larger as increasing time.
Thus the particles stay longer time around the X-type acceleration region,
and they gain more and more energy.

Figure 3 shows the ratio of the inductive electric field $|E_{y}|$
to the magnetic field $|B_{total}| = (B_x^2+B_z^2)^{1/2}$ in the X-Z plane.
This ratio becomes infinity at the X-type neutral point
because of a finite electric field $E_y$.
Roughly speaking, the electric field $E_y$
is almost uniform around the X-type reconnection region,
while the amplitude of the magnetic field $|B_{total}|$
becomes smaller at the closer vicinity of the X-type neutral point.  

We shall call the region satisfying the condition of $|E_y| \geq |B_{total}|$
in the CGS unit as the Acceleration Region (AR).
This region plays a very important role in particle acceleration and
the strength of acceleration is related to the size of the AR.

In the AR, there is no local frame that can remove $E_y$ 
by the Lorentz transformation, and
a particle is accelerated toward the Y direction.
As the outflow velocity $V_{out}$ becomes of the order of the speed of light,
the frozen-in condition $|E_{y}| = V_{out}/c |B_{total}|$ in the outer edge of
outflow region requires stronger $|E_{y}|$
and the AR grows larger.
In our case, the size of the AR is about $25\lambda \times 10 \lambda$.
It is larger than the plasma sheet thickness and
the typical spatial scale of Speiser/meandering motion.
Note that the standard ion-electron reconnection
in non-relativistic reconnection such as seen in the Earth's magnetotail
may have a small region size of $|E_y| \ge |B_{total}|$
which is much less than the electron inertia scale.
The size of AR is controlled by $V_A \sim V_{out}$.

Now we discuss the formation of the power-law energy spectrum in the AR.
First, we assume that the reconnection electric field $E_y$
is almost uniform around the X-type region,
and that particles are running very fast with the speed of light
toward Y direction.
They are continually accelerated by the strong electric field
$E_{y}$ until they are ejected from the AR.
Therefore, the acceleration efficiency may be given by
\begin{equation}
\frac{d\varepsilon}{dt} = eEc .
\end{equation}

We ignore particles running toward other directions,
because they are quickly ejected from the AR
and because they are not accelerated enough to
contribute to the high-energy part.

Second, we roughly assume that their typical life time $\tau$
around the AR is estimated
by quarter gyration of Speiser orbit in the neutral X-Y plane.
Taking into account that the Lorentz factor of
$\gamma = 1 /[1-(v/c)^2]^{1/2} $
affects the cyclotron frequency in the relativistic regime,
the loss rate with which the accelerated particles
escape from the X-type acceleration region can be given by
\begin{equation}
\frac{1}{N}\frac{dN}{dt} = -\frac{1}{\tau(\varepsilon)}
                         = -\frac{4\overline{\Omega}_z}{2\pi\gamma}
                         = -\frac{mc^2}{\varepsilon}
                         \frac{2 \overline{\Omega}_z}{\pi},
\end{equation}
where $N$ is the particle number and $\overline{\Omega}_z$ is
a cyclotron frequency by a typical value $\overline{B}_z$
of the reconnection magnetic field $B_z$. 

From these two relations, we can easily find $N \propto \varepsilon ^{-s}$,
where $s \simeq 2{\overline{B}_z}/\pi{E}$ becomes the order of 1.


\begin{figure*}
\plotone{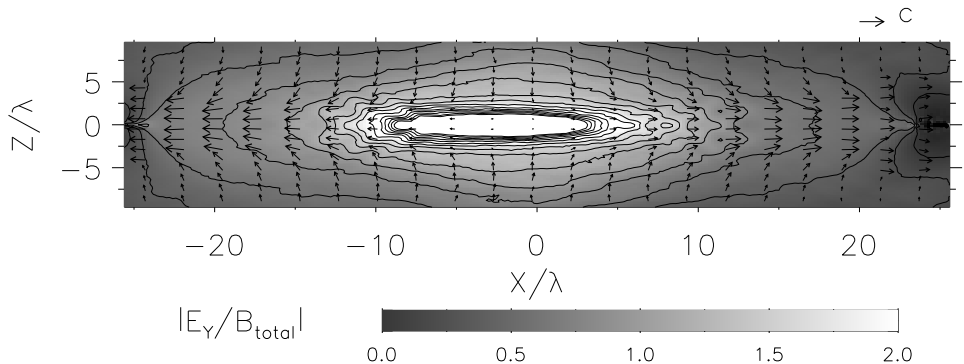}
\caption{The contour plot of the ratio of the electric field to
the magnetic field ($|E_y|/|B_{total}|$) in the simulation plane
($-25.6 \le X/\lambda \le 25.6$ and $-9.6 \le Z/\lambda \le 9.6$)
at t/$\tau_A$ = 80.8.
The vectors and the solid lines represent
the plasma flow and magnetic field lines, respectively.
\label{ratio}}
\end{figure*}

\section{Summary \& Discussions}

We have carried out simulations of magnetic reconnection
in astronomical pair-plasmas
and we found that significantly large number of non-thermal particles
are generated due to the strong inductive electric field in the AR.
Our simulation shows that particle acceleration by magnetic reconnection
can work effectively in the universe.
Moreover, the energy spectrum around the AR
resembles power-law distribution with the power-law index of 1.
This is a quite hard spectrum, which we have never yet seen.

It is highly expected that reconnection in relativistic pair-plasmas may occur
in high-energy astronomical objects,
such as a striped wind of pulsars and Active Galactic Nuclei.
The acceleration by reconnection occurs in relatively short time scale,
several tens of the \Alfven transit time
$\tau_A = \lambda / V_{A} \sim \lambda / c$,
this acceleration may be a good candidate of the origin of
non-thermal particles.

In our simulation code, we neglect any Coulomb collisions for simplicity.
This treatment is acceptable
since particles' mean free path is generally quite long in space plasmas.
They are virtually collision-less,
thus pair-annihilation is also neglected.
Considering the fact that
reconnection occur in a relatively short time scale
$\sim 10^1 \times \lambda / c$,
our assumptions may be acceptable.
Note that this code does not include any radiation losses,
such as the synchrotron and the inverse Compton losses.
In their test-particle study, \citet{schopper} have noted
that the effect of synchrotron loss is not negligible
in their high-energy situation ($10^4 \sim 10^6 mc^2$).
The effects of pair-production and pair-annihilation
are also neglected in our study.
More precise simulation should include them
but it can not be achieved using particle-in-cell code.

We have also discussed the formation of power-law spectrum
around the X-type region
and our simulation result is well described by $N \propto \varepsilon ^{-s}$,
where the index $s$ is about the unity.
Strictly speaking,
it is unable to discuss the classical Speiser/meandering orbit
in the vicinity of the X-type region where $|E_y| \ge |B_{total}|$,
since the characteristic of Lorentz transformation changes.
Particle orbits are still similar to Speiser/meandering orbit,
but they are driven by $E_y$ and resonant with $E_y$.
We are still in the process of searching
analytical solutions for the acceleration orbit,
but the essential point of our model is as follows:
the more strongly particles are accelerated in the Y direction,
the longer time they stay in the AR,
due to the relativistic effect of cyclotron-motion.

In our run, the growth of the maximum energy $\varepsilon_{max}$
seems to be confined by the periodic boundary.
Simulation run with larger simulation box
may produce more energy than $38mc^2$.
The multiple reconnections may be also important for
producing high-energy particles.
If two or more reconnection regions are formed in the plasma sheet,
ejected particles travel into another AR so that they can be more accelerated.

We have also performed simulation runs
with different \Alfven speed parameters.
We have recognized remarkable number of non-thermal particles in every case.
Together with dependency of reconnection behavior
on the plasma sheet thickness,
we will report the full story in the coming paper.

\acknowledgments
The authors are grateful to T. Terasawa for fruitful discussions.

\end{document}